\documentclass{article}
\usepackage{graphicx}
\usepackage{graphics}
\usepackage{subfigure}
\usepackage{booktabs}
\usepackage{authblk}

\usepackage[numbers,square]{natbib}

\usepackage{latexsym,amssymb,enumerate,amsmath,bm}
\usepackage{psfrag}

\newcommand{\beq}{\begin{equation}}
\newcommand{\eeq}{\end{equation}}

\newcommand{\tc}{\tau_c}

\newcommand{\vs}{\left( \begin{array}{c}}
\newcommand{\ve}{\end{array}\right)}

\title{Force-clamp experiments reveal the free energy profile 
	and diffusion coefficient of the collapse of proteins}

\author[1]{H. Lannon}
\author[2]{E. Vanden-Eijnden}
\author[1]{J. Brujic}
\affil[1]{Department of Physics and Center for Soft Matter Research, New York University, 4 Washington Place, New York, NY, 10003, USA}
\affil[2]{Courant Institute of Mathematical Sciences, New York University, 251 Mercer St, New York, NY, 10012, USA}

\date{\today}
\begin{document}
\maketitle

\begin{abstract}
  We present force-clamp data on the collapse of ubiquitin
  polyproteins in response to a quench in the force. These
  nonequilibrium trajectories are analyzed using a general method
  based on a diffusive assumption of the end-to-end length to
  reconstruct a downhill free energy profile at $5$pN and an energy
  plateau at $10$pN with a slow diffusion coefficient on the order
  of~$100$nm$^2$s$^{-1}$. The shape of the free energy and its linear
  scaling with the protein length give validity to a physical model for
  the collapse. However, the length independent diffusion coefficient suggests that internal rather than viscous 
  friction dominates and thermal noise is needed to capture the variability in the measured times to collapse.

\end{abstract}

%\keywords{activated processes; reaction
  %coordinate}%Use showkeys class option if keyword

By measuring the end-to-end length of proteins and RNA in response to
force perturbations, single molecule experiments open a window into
the complex dynamics of these molecules on their multi-dimensional
energy potentials~\cite{Fernandez2004, Bustamante1994,
  Woodside2006}. For example, a protein is unfolded by the application
of a constant pulling force, while quenching the force to a low value
triggers the hydrophobic collapse of the molecule~\cite{Fernandez2004,
  Garcia-Manyes2007}. This dynamical collapse has been modeled as a
one-dimensional diffusion of the measured end-to-end length on a free
energy profile in the case of protein monomers~\cite{Berkovich2010}
and RNA molecules~\cite{Hyeon2009}. By contrast, dynamics
in degrees of freedom hidden from the experiment were thought to
govern the large diversity in the end-to-end length of trajectories
visited by collapsing polyproteins~\cite{Walther2007}. Whether the
experimental distribution of trajectories can be described by simple
diffusion along the measured reaction coordinate or requires multiple
dimensions remains an open question that requires novel analysis
tools.

This question is non-trivial because the collapsing traces are
out-of-equilibrium and standard techniques to reconstruct the free
energy profile based on the Jarzynski
equality~\cite{Jarzynski1997} or Crook's fluctuation
theorem~\cite{Crooks1998} are not applicable to the force
quench experimental protocol. Indeed, these techniques rely on knowing
the statistics of the work exerted on the system~\cite{Hummer2001,
  Liphardt2002,Collin2005,Minh2008}.  In force-clamp experiments this
work is concentrated in the brief time it takes to quench the force
($\sim$50ms), which would require a prohibitively large pool of data
to ensure the statistical accuracy of the free energy estimator based
on the Jarzynski
equality~\cite{Hummer2001b,Oberhofer2005,Oberhofer2009,Minh2009}.  A
second difficulty is that the free energy alone is not sufficient to
describe the dynamics of the collapse. If the collapse can be
described by an overdamped Langevin equation for the end-to-end length
of the protein~\cite{Berkovich2010}, then a diffusion coefficient must
be estimated besides the free energy~\cite{Best2010,Schlagberger2006}.

Here we introduce an analysis method to reconstruct the free energy
profile~\cite{Zhang2011} directly from the collapse trajectories of
ubiquitin polyproteins, assuming diffusive dynamics. By
reconstructing the free energy for polyprotein chains with varying
numbers of protein domains, we quantify to what extent the collapse
mechanism is cooperative between the
domains~\cite{Fernandez2004b}. Moreover, the observation that
increasing the quench force slows down the collapse process
\cite{Garcia-Manyes2007,Fernandez2004} is explained in terms of the
shape of the reconstructed free energy landscape, which in turn tests
the Bell model \cite{Bell1978} with no adjustable parameters. We then present
an extension to the approach that offers the first measurement of an
effective diffusion coefficient of a collapsing polypeptide and tests
its constancy along the measured reaction coordinate. Finally, we
propose a microscopic origin for the observed collapse in terms of the worm like chain and `expanding sausage' models~\cite{Hyeon2009}. 

\begin{figure}[htbp]
\begin{center}
\centerline{\includegraphics[width=0.9\textwidth]{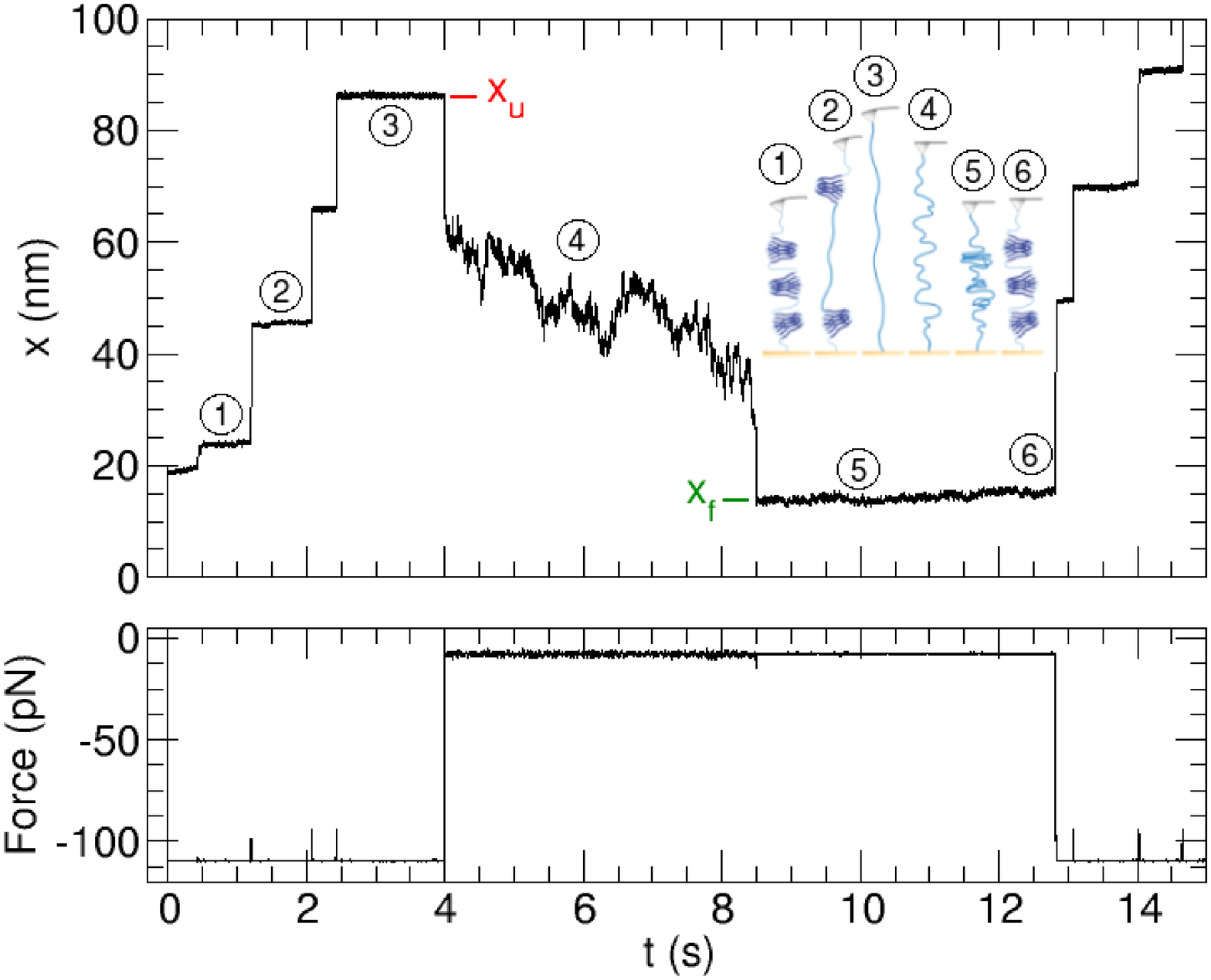}}
\caption{A typical force-clamp trajectory of the unfolding and
  refolding of a polyubiquitin chain with $N_d=3$ domains. A second
  pull to $110$pN reveals a staircase as a signature that the protein
  domains refold.}
\label{fig:fulltrace}
\end{center}
\end{figure}

We use Atomic Force Microscopy (AFM) in the force-clamp mode to follow
the unfolding and refolding trajectories of ubiquitin polyproteins
under a constant stretching force, as shown in the example in
Fig.~\ref{fig:fulltrace}. Exposing a mechanically stable protein to a
high pulling force of $110$pN leads to the stepwise unfolding and
extension of each of the three protein domains in the polypeptide
chain. Subsequently, quenching the force to a lower value of $10$pN
triggers the collapse of the whole protein from a fully extended state 
back to a collapsed state with the same end-to-end length as the
folded protein. Previous experiments have shown that the final state
of the collapse process does not lead to a mechanically stable folded
protein, but a random compact globule that forms native contacts over
time \cite{Garcia-Manyes2009}. A second pull on the same protein at
$110$pN leads to a second unfolding, as shown in the trajectory. Here
we analyze only those trajectories that exhibit a minimum of three
steps of $\sim$20nm in the initial staircase as a signature of the
extension of individual ubiquitin domains upon unfolding, as well as a
second staircase to signify refolding. The question is then to
understand the mechanism of the collapse dynamics from many recordings
($n_{tot}\sim100$) of these trajectories.

Theoretically, if we denote by $x$ the end-to-end length, the
overdamped Langevin equation reads
\begin{equation}
  \label{eq:1}
  \dot x = -\beta D G'(x) + \sqrt{2D}\, \eta(t) 
\end{equation}
where $\beta = 1/(k_B T)$, $\eta(t)$ is a white-noise term accounting
for thermal effects, $G(x)$ is the equilibrium free energy profile and
$D$ is the diffusion coefficient which we assume to be constant (this
assumption is validated below).  Both $G(x)$ and $D$, or the friction
coefficient $\gamma$ since $D= k_B T/\gamma$, can be estimated from
the collapsing traces using the techniques introduced
in~\cite{Zhang2011}.

Let us consider the free energy first. The procedure to calculate
$G(x)$ from the collapsing traces is to cut out pieces of trajectories
from the moment the force is quenched at the unfolded length,~$x_u$,
until the moment they first reach the folded length at low
force,~$x_{\!f}$. This allows us to estimate via binning a
nonequilibrium stationary probability density $\rho(x)$ of many
collapsing trajectories, and relate $G(x)$ to it for
$x\in[x_{\!f},x_u]$ as follows:
\begin{equation}
  \label{eq:2}
  G(x) = - k_B T \ln \rho(x) - k_BT\rho'(x_{\!f}) 
  \int_{x}^{x_u} dx'/\rho(x')
\end{equation}
where $\rho'(x_{\!f})$ denotes the derivative of $\rho(x)$ estimated
at~$x_{\!f}$. Note that this formula is different from the standard
$G(x) = - k_B T \ln \rho_e(x)$, where $\rho_e(x)$ is the equilibrium
probability density function. The nonequilibrium $\rho(x)$ requires an
additional term besides $- k_B T \ln \rho(x)$ in~\eqref{eq:2} to
relate it to $G(x)$. This extra term corrects for the fact that
$\rho(x)$ is biased towards values of~$x$ that are closer to~$x_u$,
where the trajectories are initiated by the protocol. For a detailed
derivation of~\eqref{eq:2} we refer the reader to~\cite{Zhang2011},
where this formula is also compared to Bayesian inference
methods~\cite{Best2010, Best2011, Pokern2009}. Using Eq.~\eqref{eq:2} is advantageous
because the chronological order in which the data is acquired does not
play a role in the binning procedure, which implies that the time
resolution of the instrument ($\sim5$ms) has no impact on the
resulting landscape.

\begin{figure}[t]
\begin{center}
\centerline{\includegraphics[width=0.9\textwidth]{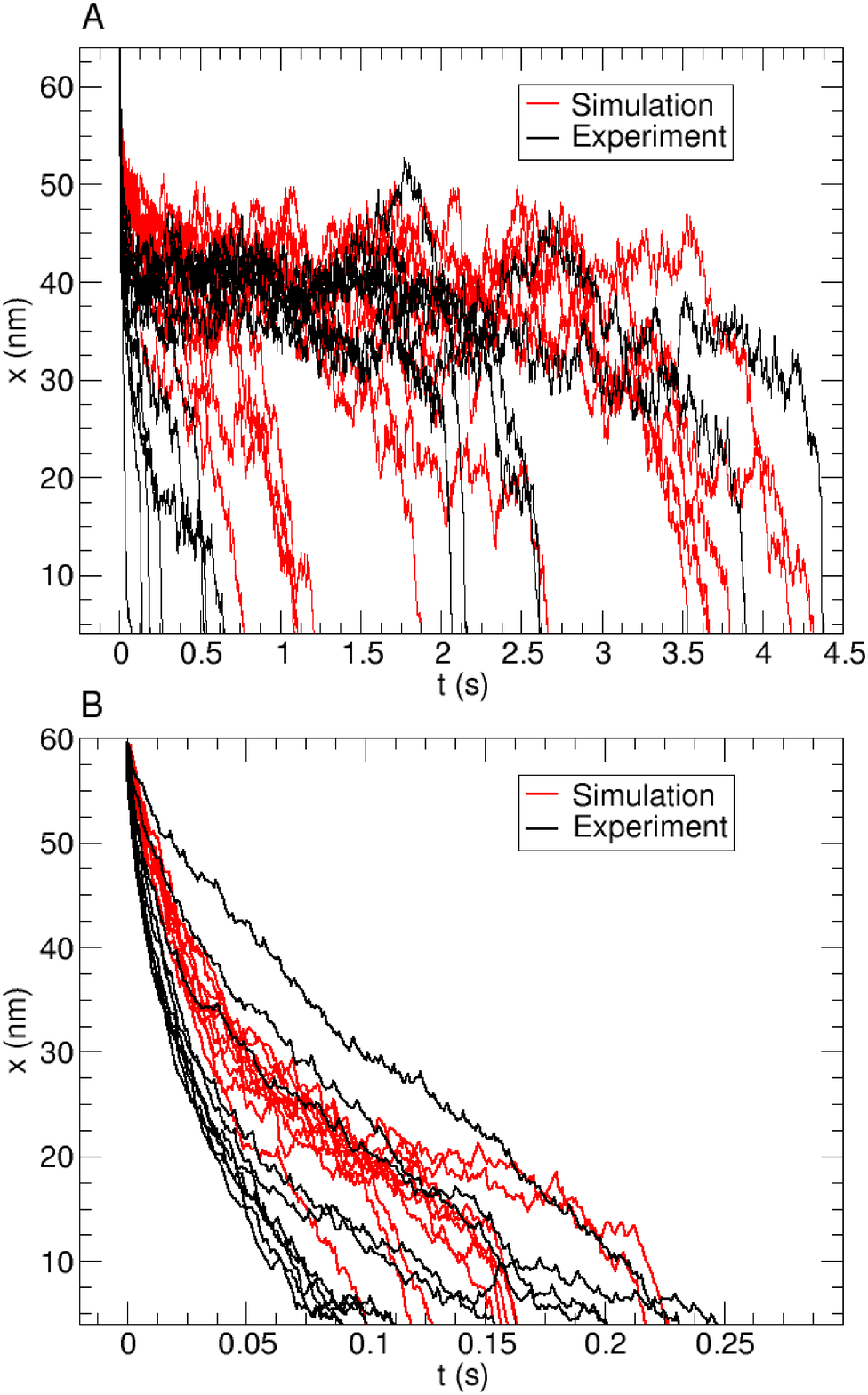}}
\caption{Collapsing trajectories are grouped by their total length
  ($N_d=3$) and aligned at the time of the force quench to $10$pN in
  (A) and $5$pN in (B). The experimental trajectories are compared
  with those generated by simulations of diffusive dynamics on the
  reconstructed free energy profiles.}
\label{fig:collapse}
\end{center}
\end{figure}

\begin{figure}[htbp]
\begin{center}
\centerline{\includegraphics[width=0.9\textwidth]{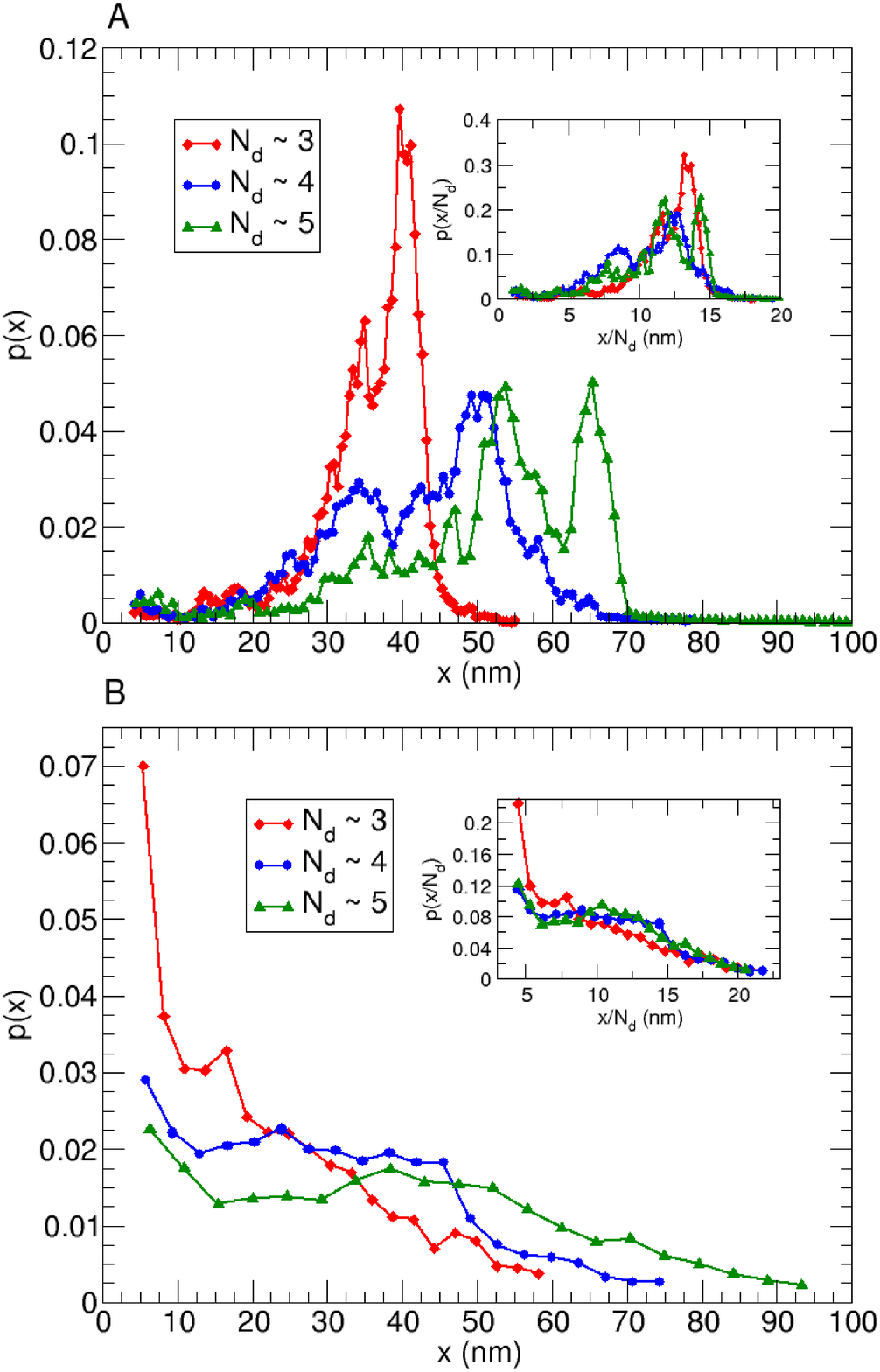}}
\caption{The nonequilibrium distribution $\rho(x)$ for each $N_d$
  collected at a force quench of $10$pN in (A) and $5$pN in (B). The
  linear rescaling by $N_d$ is shown in the inset, which indicates a
  cooperative mechanism for the collapse.}
\label{fig:densities}
\end{center}
\end{figure}

Next we apply~\eqref{eq:2} to analyze force-clamp trajectories, such
as the one shown in Figure~\ref{fig:fulltrace}. Since the length of
the polyprotein chain and the polypeptide linker to the surface vary
from one experiment to the next, we compare all trajectories in terms
of the total length of the collapse~$L_{tot}=x_u-x_{\!f}$. We find
that~$L_{tot}$ clusters in increments of a monomer ubiquitin length of
$\sim20$nm with a standard deviation of $\sim6$nm. We therefore group
the clusters of similar collapse lengths and estimate the number of
domains in the polyprotein chain as $N_d=L_{tot}/20$nm to the nearest
integer. Setting the lowest value of $L_{tot}$ within a group of a
given $N_d$ to be $x_u$ at time zero and $x_{\!f}$ to $4$nm~\cite{Oesterhelt2000} as the
folded length of the protein from the protein data bank, leads to the alignments of trajectories
shown in Figs.~\ref{fig:collapse}A and~\ref{fig:collapse}B for the
$10$pN and $5$pN force quench, respectively, in the group of $N_d=3$.
Analyzing trajectories in groups segregated by~$N_d$, we measure the
non-equilibrium distribution $\rho(x)$ of the end-to-end length for
each~$N_d$, as shown in Fig.~\ref{fig:densities}. We find that they
approximately scale linearly with~$N_d$ at both forces, as shown in
the insets. At a quench force of $10$pN, the extended polypeptides
often plateau at $\sim70\%$ of the contour length before their final
collapse. Lowering the quench force to $5$pN reveals faster collapse
trajectories that visit all end-to-end lengths with a similar
probability. 
%Note that in the $5$pN case we exclude from the present
%analysis $\sim30\%$ of trajectories that took between 2 and 15 seconds
%to collapse. Given that they did not occur in the 10 trajectories in
%the $N_d=3$ group and became more prevalent in the collapse of very
%long chains, we postulate that another mechanism beyond the scope of
%this study may occur.

Using the observed distributions, we then obtain $G_{N_d}(x)$, the free
energy of a polyprotein of~$N_d$ units, and collapse these different
profiles onto one another using the rescaling
\begin{equation}
  \label{eq:4}
  G(x+x_{\!f}) \equiv G_{N_d=1}(x+x_{\!f}) = 
  \frac1{N_d}G_{N_d}\left(N_d(x +x_{\!f})\right)
\end{equation}
%
%This rescaling shown in Fig.~\ref{fig:fereconstruct}A is consistent
%with a polymer model for the recoil of an entropic spring, such as the
%wormlike chain model \cite{Bustamante1994}, which one expects to observe in
%the free energy profile at large~$x$. Moreover, this scaling also
%extends to low values of $x$ closer to $x_{\!f}$, where hydrophobic
%interactions along the chain play an important role. 
This cooperativity between the domains is inconsistent with previously
proposed models for the stochastic refolding of individual domains
\cite{Best2005} or the aggregation of the unfolded domains
\cite{Wright2005}. Instead, our result in (\ref{eq:4}) suggests a global collapse of the polypeptide chain due to the attraction
between hydrophobic residues that do not directly lead to folding
\cite{Garcia-Manyes2009,Xia2011,Chandler2005}.

\begin{figure}[htbp]
\begin{center}
\centerline{\includegraphics[width=0.9\textwidth]{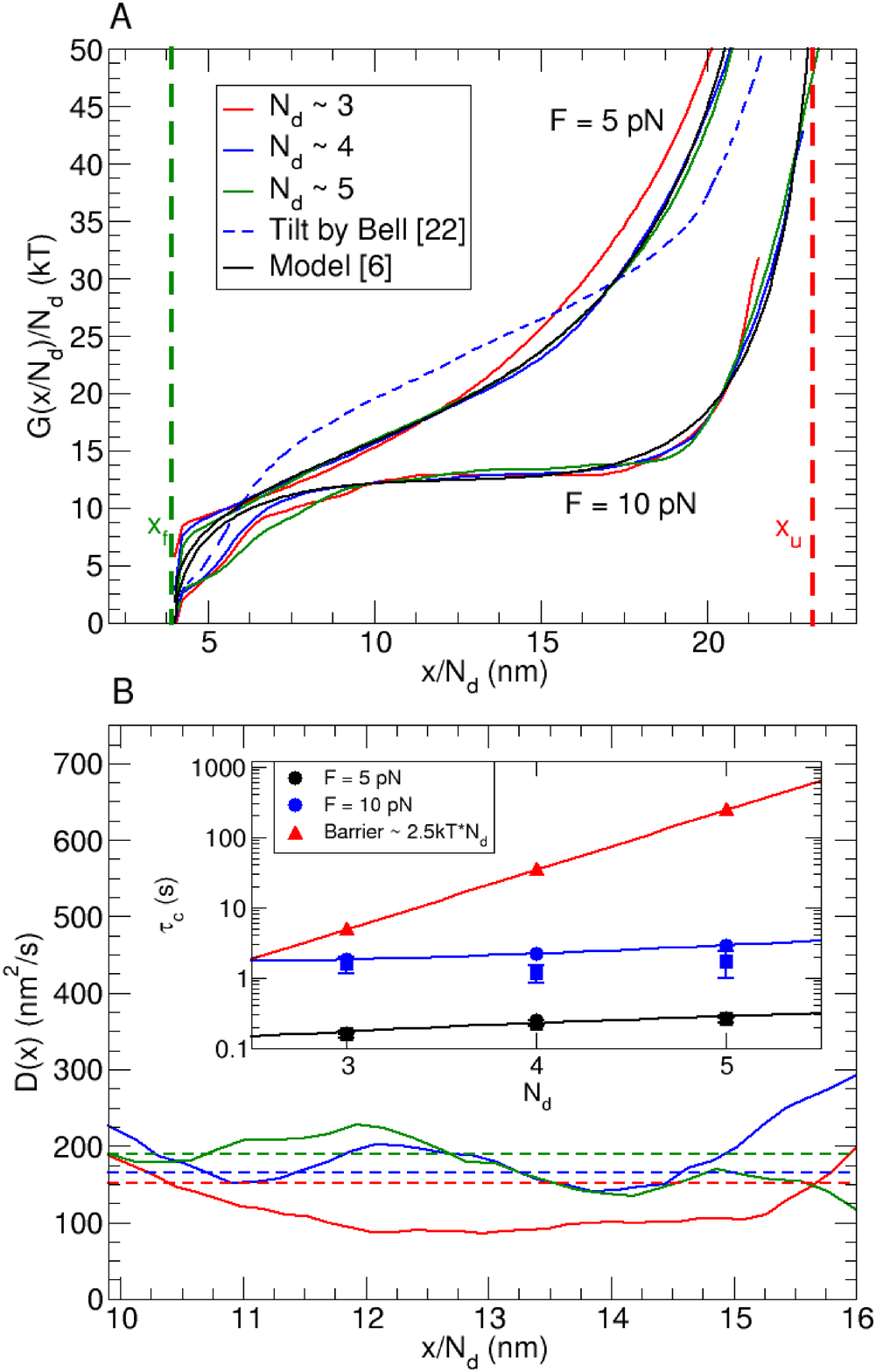}}
\caption{(A)Experimental free energy profiles as a function of the
  end-to-end length, rescaled by $N_d$. (B) Diffusion coefficients
  derived from Eq. (5) at $10$pN (solid lines) compare well with those 
  estimated from the free energy reconstruction (dashed lines).
  The inset shows $\tau_c$ dependence on $N_d$ (squares), consistent 
  with simulated data (circles)}
\label{fig:fereconstruct}
\end{center}
\end{figure}

The shape of the resulting free energy profile $G(x)$ per ubiquitin
monomer in Fig. 4A at $10$pN is interesting because of the absence of
a barrier: the experimental collapse corresponds to a diffusive slide
on a plateau in the free energy that accelerates as the end-to-end
length reaches a value~$\sim5$nm away from $x_{\!f}$. Lowering the
force to $5pN$ eliminates the plateau landscape and promotes a
downhill collapse that is limited by friction alone, which is roughly
consistent with the prediction of the tilt by the Bell model, also
shown in Fig. 4A. Similar features of ubiquitin monomer trajectories
under a quench force of $10$pN were interpreted in terms of a physical model
that predicts a free energy profile with a barrier of $2.5k_BT$~\cite{Berkovich2010}. Since
tilting the profiles in Fig.~\ref{fig:fereconstruct}A by the Bell
model~\cite{Bell1978} to just $13$pN leads to a barrier to collapse of
the same height, this small difference in the quench force could
explain the observed change in the profile. However, the functional
form of the landscape proposed in~\cite{Berkovich2010} does not fit
the free energy profiles accurately due to its propensity to form
barriers over a wide range of quench forces.

A better fit is achieved using the physical model
proposed for the collapse of RNA molecules in~\cite{Hyeon2009}, which
is based on the sum of the entropic worm like chain model, the work done on
the protein and the enthalpic `expanding sausage' model for
polypeptide collapse~\cite{degennes1985}:
\begin{equation}
\label{eq:sausage}
\begin{aligned}
& G(x) = \frac{2k_B T}{\xi^2} \sqrt{\pi \Omega (x-x_{\!f})} - F \, (x-x_{\!f}) \\
	&\quad + k_BT\frac{L_c}{l_P}\int_0^{\frac{x-x_{\!f}}{L_c-x_{\!f}}} 
        \left(\frac1{4(1-y)^2}-\frac14 + y\right) dy
	\end{aligned}
\end{equation} 
Here $F$ is the applied force, $L_c$ and $l_P$ are the contour and
persistence lengths of the extended protein, respectively, $\Omega$ is
the volume of the sausage, and $\xi$ is the size of a globule inside
the sausage~\cite{degennes1985}.
The adjustable parameters in Eq.~(\ref{eq:sausage}) are $l_P$, $L_c$, and the 
ratio $\sqrt{\Omega}/\xi^2$.  Fits to $G(x)$ in Fig.~\ref{fig:fereconstruct} give
$L_c=26$nm, predicted by the size of a ubiquitin monomer ($76$~residues~$\times~0.36=27.4$nm)~\cite{Oesterhelt2000},
$l_P=0.82$nm at $5$pN and $1.45$nm at $10$pN, in agreement with chain
stiffening along the backbone due to intramolecular
interactions~\cite{Walther2007}. To obtain the values of $\Omega$ and $\xi$ 
from their ratio given by the fit, we assume that the size of the individual
monomers in the sausage is $l_P$. 
%consistent with our
%observation that this parameter varies with the force as the chain
%stiffens. 
This implies that the number $N$ of these monomers must be $N =
L_c/l_P$. Following de Gennes' argument, we then set $\xi= l_P
\sqrt{g}$ and $\Omega = L_c \pi \xi^2 = L_c \pi l_P^2g$, where $g$ is
the number of monomers inside each globule and 
becomes the fit parameter that replaces the ratio. 
Fits to $G(x)$ in
Fig.~\ref{fig:fereconstruct} thus yield $\xi=2.6$nm at $5$pN and $2.7$nm at $10$pN,
in rough agreement with the value $\xi=2$nm estimated for the
hydrophobic collapse~\cite{Thirumalai1997}, and $\Omega=203.62$nm$^3$
at $5$pN and $373.90$nm$^3$ at $10$pN. Note that the above argument
does not affect the quality of the fits, simply it gives an
interpretation of the parameters in Eq~\eqref{eq:sausage} that
indicates that the microscopic packing of blobs inside the initial
sausage is different for the two quench forces. Note also that the
functional form of this free energy is consistent with the scaling
with $N_d$ in Eq.~(\ref{eq:4}) since the volume of a polyprotein with
$N_d$ domains is $N_d\Omega$ and its contour length $N_dL_c$ while all
the other parameters in Eq.~(\ref{eq:sausage}) are unaffected
by~$N_d$. Altogether, these results give a quantitative validation of
the physical model underlying the collapse.
%$g = ...$ at $5$pN and $...$nm at
%$10$pN. In turn this gives 

The collapsing traces can also be used to calculate the diffusion
coefficient~$D(x)$ on the reconstructed landscape and thereby verify
our assumption that it is constant, $D(x) \approx D$.  The idea is to
replace $x_{\!f}$ by any $ x \in[x_{\!f},x_u]$ in the procedure,
i.e. cut the traces from $x_u$ till the first moment they reach $x$
and recalculate their non-equilibrium probability density $\rho$. The
probability flux of these traces through the end-point $x$ can be expressed in
two ways: it is given by $D(x)\rho'(x)$, and it can also be
estimated directly as $1/\tau_c(x)$, where $\tau_c(x)$ is the average time
it takes them to collapse from $x_u$ to $x$. Equating these two
expressions and solving for $D(x)$ gives
\begin{equation}
  \label{eq:3}
  D(x) = 1/(\tc(x)\rho'(x))
\end{equation}
This estimator for $D(x)$ is new and it has the advantage over the
standard one using quadratic variation of the
trajectory~\cite{Best2010} that it is insensitive to the time
resolution of the instrument. Because of the small number of traces
per $N_d$ per force ($\sim 15$), the estimate for $D(x)$ is
accurate over the plateau regime in the end-to-end length in the data
set at $10$pN and not in the drift dominated parts of
the landscape. The results obtained for polyproteins with
different~$N_d$ in Fig.~4B are in good agreement with each other,
within the experimental error, and show that the diffusion coefficient
is roughly constant as a function of $x$, consistent with the
assumption made in Eq.~\eqref{eq:1}. This is a surprising result
because the `expanding sausage' model predicts a $1/x$ scaling of
$D(x)$ due to an increase in the viscous friction as the molecule
collapses to a blob of a growing radius.  By contrast, here the protein dynamics is
governed by internal rather than solvent friction, which agrees
with recent single molecule experiments that show an independent
friction with the end-to-end length of a folding protein~\cite{Soranno2012,
 Cellmer2008}. Notice also that the average value of $170$nm$^2$/s is
orders of magnitude smaller than the typical vibrational modes of a
protein \cite{Go1983}. This indicates that the projection of all the
degrees of freedom of the molecule onto a single reaction coordinate
manifests itself as a very slow diffusion. It is likely that many
local barriers in other degrees of freedom (associated with the
formation of e.g. loops or helices at the same end-to-end length) can
be mimicked by an effective diffusion constant. 
%Next we show that this
%surprising result is in perfect accordance with the
%experimental data.

To verify our results, we generate artificial traces using Eq.~(1)
with the estimated $G(x)$ and $D$ and show that they are in excellent
agreement with experimental traces in Fig.~2. In addition, the fact
that traces generated using $D$ derived at $10$pN reproduce the spread
of times to collapse and the noise fluctuations in the experimental
traces at~$5$pN suggests that $D$ does not change with the quench
force. We estimate that $\sim70\%$ of experimental trajectories are consistent with the 1-D
diffusive model, while the outliers do not agree with the synthetic distribution of collapse times. Such trajectories have been observed previously~\cite{Walther2007, Garcia-Manyes2009} and they highlight the importance of other degrees of freedom. Nevertheless, the simulated and the experimental average times
to collapse $\tau_c$ agree very well at both quench forces and for all
$N_d$, as shown in the inset in Fig.~\ref{fig:fereconstruct}B. By contrast, a linear scaling with $N_d$ of a barrier-limited $G(x)$~\cite{Berkovich2010} would lead to a much steeper dependence of $\tau_c$ with $N_d$, which is inconsistent with our and other published polyprotein data~\cite{Fernandez2004}.
 
This general non-equilibrium method to analyse single molecule
trajectories has allowed us to reconstruct free energy profiles, assess the dynamics along the measured reaction coordinate and thus
postulate a physical model for the collapse of ubiquitin
proteins. This technique paves the path for a mechanistic approach to
many complex problems, such as protein folding.    

We would like to acknowledge Jennifer Haghpanah and Jin Montclare for the
expression of ubiquitin polyproteins, as well as Alexander Grosberg for useful discussions. J.~B. holds a Career Award at the Scientific Interface from the Burroughs Wellcome Fund and was supported in part by New York University Materials Research Science and Engineering Center Award DMR-0820341 and a Career Award 0955621.

%%%%%%%%%%%%%%%%%%%%%%%%%%%%%%%%%%%%%%%%
%
%        B I B L I O G R A P H Y

\bibliographystyle{unsrtnat}
\bibliography{bibl}

\end{document}